\documentclass[aps,amsfonts,amsmath,prl,tightenlines,twocolumn,showpacs,nofootinbib]{revtex4}
\usepackage{epsf}

\def\be{\begin{equation}}
\def\ee{\end{equation}}
\def\beq{\begin{equation}}
\def\eeq{\end{equation}}
\def\bea{\begin{eqnarray}}
\def\eea{\end{eqnarray}}
\def\bml{\begin{subequations}}
\def\blea{\bml\begin{eqnarray}}
\def\elea{\end{eqnarray}\end{subequations}}

\begin{document}

\title{Scaling of cosmic string loops}

\author{Vitaly Vanchurin}
\email{vitaly@cosmos.phy.tufts.edu}
\author{Ken D. Olum}
\email{kdo@cosmos.phy.tufts.edu}
\author{Alexander Vilenkin}
\email{vilenkin@cosmos.phy.tufts.edu}
\affiliation{Institute of
Cosmology, Department of Physics and Astronomy, Tufts University,
Medford, MA  02155}

\begin{abstract}

We study the spectrum of loops as a part of a complete network of
cosmic strings in flat spacetime.  After a long transient regime,
characterized by production of small loops at the scale of the initial
conditions, it appears that a true scaling regime takes over. In this
final regime the characteristic length of loops scales as $0.1 t$, in
contrast to earlier simulations which found tiny loops. We expect the
expanding-universe behavior to be qualitatively similar. The large
loop sizes have important cosmological implications. In particular,
the nucleosynthesis bound becomes $G\mu \lesssim 10^{-7}$, much
tighter than before.

\end{abstract}


\maketitle

Cosmic strings could be formed as linear defects at symmetry breaking
phase transitions in the early universe \cite{Kibble}. Alternatively,
they could arise as fundamental or $D$-strings at the end of brane
inflation \cite{Tye,Dvali,Polchinski}.
Strings can produce a variety of observational effects, and searches
are now underway for their signatures in gravitational lensing, CMB
anisotropies, and gravitational wave background. A good theoretical
understanding of string networks is crucial for interpreting the
results of these searches. However, despite much effort, 
the evolution of cosmic strings is not yet fully understood.

An evolving string network consists of two components: long strings
and sub-horizon closed loops. It is fairly well established that long
strings exhibit scaling behavior: both the average distance between
the strings $d(t)$ and the coherence (or persistence) length $\xi(t)$ 
scale with the
cosmic horizon,
\beq
d(t)\sim\xi(t)\sim t.
\label{dxi}
\eeq

In the early work on cosmic strings, it was expected
\cite{AV81,Kibble85} that the typical length of closed loops scales in
a similar manner, $l(t)\sim t$.  The first numerical
simulations of string evolution seemed to support this scenario
\cite{Turok}.  However, later, high-resolution simulations
\cite{BB,AS} showed that
the loop sizes were actually
much smaller than the horizon and gave no evidence for
scaling. On the contrary, the typical loop size remained nearly
constant and close to the resolution limit of the simulations. The
simulations also revealed that long strings had a significant
substructure, with short-wavelength wiggles all the way down to the
resolution limit.

The standard scenario of string evolution that has emerged from these
findings assumes that the typical size of loops $l(t)$ is set by
the scale of the smallest wiggles, which is in turn determined by
damping due to gravitational radiation \cite{BB91}. The typical loop
size is then given by 
\beq 
l(t)\sim \alpha t
\label{lmin}
\eeq
with \cite{OSV} $\alpha\sim(G\mu)^\gamma$, 
where $G$ is Newton's constant and $\mu$ is the mass per unit length
of string. With plausible assumptions about the spectrum of wiggles,
the power index $\gamma$ is in the range $2\leq \gamma \leq 3$. The
observational bound on the string mass parameter $G\mu$ is
$G\mu < 10^{-6}$, and the corresponding bound on $\alpha$ is
$\alpha < 10^{-12}$, indicating that the loops are extremely
small.

A more radical string evolution scenario, proposed in
\cite{Hindmarsh}, suggests that the loops are even smaller. It claims
that strings lose most of their energy by direct emission of
microscopic loops of size not much greater than the string thickness.
This idea was not confirmed in other simulations \cite{MS,OBP}.  It is
hard to tell which, if any, of these scenarios is correct, since the
loop sizes they suggest are well beyond the resolution limits of
current simulations.

To address this issue, we developed in \cite{VOV} a flat-space string
simulation which does not suffer from the problem of smallest
resolution scale. This simulation uses functional forms for the string
positions and is exact to the limits of computer arithmetics. In
Ref. \cite{VOV} we used our simulation to show that the spectrum of
wiggles on long strings scales with time, even in the absence of
gravitational damping. The spectrum has a universal form; it is peaked
at $l\sim 0.3 t$ and declines slowly towards smaller scales.

In the present paper we report on the study of the closed loop
component of the simulation. We find that loops are produced in a wide
range of sizes, with most of the string length going into relatively
large loops of size $l \sim 0.1 t$, comparable to the inter-string
distance.  This scaling behavior is established only after a long
transient regime, characterized by copious production of tiny loops
whose size is set by the scale of the initial string network. We
believe that it was this transient regime that was observed in earlier
simulations.

The numerical results presented here are based on the computer
simulation described in \cite{VOV}. In order to be able to study the
evolution at late times we periodically increase the simulation volume
as described in \cite{VOV}. Specifically, we create 8 identical
replicas of the simulation box and glue them together to form a box
twice the size of the original one. The resulting unphysical
correlations on super-horizon scales are removed by allowing the
string reconnection probability to be $p<1$. We used the value of
$p=0.5$ which maximizes the rate of decay of the
correlations.\footnote{To minimize disturbance to the string network
we do not use the smoothing procedure of \cite{VOV}, and we keep the
reconnection probability always 0.5, rather than switching between 0.5
and 1. Apart from making the time dependence smoother, this has no
significant effect on the results.}  We have verified that simulations
with $p = 0.2$ and $0.8$ give similar results.

Until recently, there was a widespread opinion that all values of $p$
other than $p=1$ are unphysical and therefore of little
interest. However, it was pointed out in \cite{Dvali,Tye1,Jones} that
for fundamental and $D$-strings $p$ generally differs from 1, and may
even be $\ll 1$. The dependence of network evolution on the value of
$p$ is an important problem, but we will not try to address it here. 
It will require more extensive simulations, since the evolution is
very slow for small $p$.

To start our simulation we generate four different Vachaspati-Vilenkin
initial conditions \cite{VV} and overlay them in the same simulation
box. We displace the four realizations relative to each other so that
equivalent lattice points lie on the corners of a tetrahedron of
height 0.5.  In previous simulation work we found that the correlation
length $\xi(t)$ is larger than the inter-string distance $d(t)$ by a
factor of about 2.6 in the scaling regime, while in the initial
conditions this factor is only about 1.3.  By overlaying four initial
conditions, we decrease the initial inter-string distance by a factor
2, while not changing the correlation length, so the initial
conditions have properties more similar to a scaling regime.

The maximum initial box size with four interlaced initial conditions
that we can simulate on a single processor with 3GB of memory is 50,
which becomes 800 by time 1000, after 4 doublings of the box size. The
results presented here are the averages of 25 simulations.  The
expansion of the simulation volumes takes place at typical times 33,
93, 259, and 924.  

\begin{figure}
\vskip -10pt
\moveleft 1in\hbox{\epsfxsize=5.5in\epsfbox{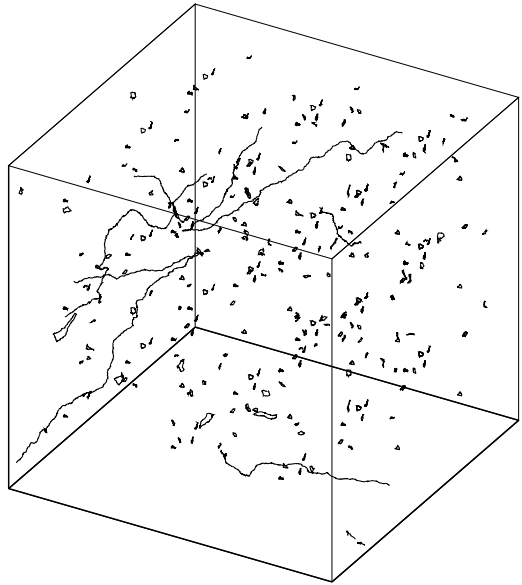}}
\vskip -10pt
\caption{A $150^3$ section of the network at time 800.  Loops of sizes
less than 10 are not shown.}
\label{fig:network}
\end{figure}

A snapshot of a cubic section of the network at time 800
can be seen in Fig.\ \ref{fig:network}.
There are some long strings crossing the cube, many small loops,
and a few loops of sizes up to 60, comparable to the average
inter-string separation at that time.

\begin{figure}
\begin{center}
\leavevmode\epsfxsize=3.0in\epsfbox{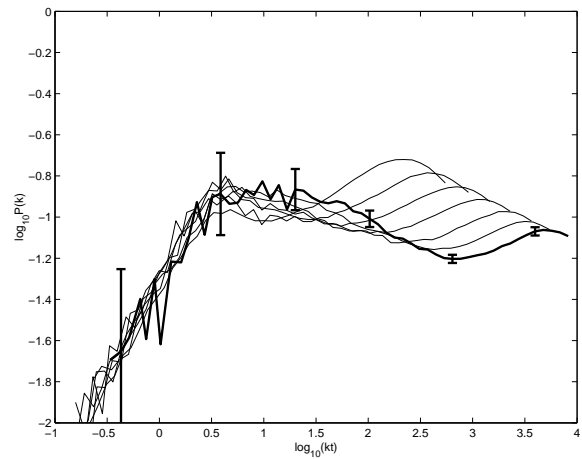}
\end{center}
\caption{Evolution of the power spectrum $P(kt)$ vs.\ $kt$ at times:
60, 100, 160, 250, 400, 600, 900 (bold).  Error bars are the one sigma
run-to-run variation.}
\label{fig:spectrum-kt}
\end{figure}

The spectrum of wiggles, as defined in \cite{VOV}, is shown in Fig.\
\ref{fig:spectrum-kt}
as a function of $kt$. There are two peaks present on the plot: one
scaling and one non-scaling. The non-scaling peak slowly decays and
moves to large values of $kt$, but stays roughly at the same value of
$k$ that corresponds to the initial correlation length. In contrast,
the scaling peak remains at almost the same value of $kt$. Note that
in \cite{VOV} the non-scaling peak was completely eliminated by
smoothing.  We have not used smoothing here, because it potentially
distorts the spectrum of small loops and because the jumps it
introduces make it difficult to see trends in the loop production
spectrum.

We characterize the rate of loop production by the function $n(l,
t)$ --- the number of loops produced per unit loop length per unit
volume of the network per unit time.  In a scaling network, the number
of loops with sizes between $l$ and $l+dl$ produced in a volume $L^3$
evolving from time $t$ to $t+dt$ is the same as the number with sizes
between $2l$ and $2l+2dl$ produced in a volume $8L^3$ evolving from
time $2t$ to $2t+2dt$.  Thus, for scaling, \be\label{eqn:f} n(l, t) =
t^{-5} f(x), \ee where $f$ can be any function of $x = l/t$.

In a cosmological string network, infinite strings self-intersect and
produce loops. These loops can ether reconnect with other strings,
fragment further by self-intersections, or oscillate without
self-intersections.  But in a simulation, all strings are closed, so
we need some definition of the point at which a loop is considered to
have been produced.  We proceed as follows.

First, we say that a loop is a survivor if neither it, nor any
fragment produced from it, rejoins any other string.  To conserve
computer memory, we do not allow strings shorter than a minimum length
$\kappa t$ to rejoin the network, so any string smaller than $\kappa
t$ is automatically a survivor.  We have verified in \cite{VOV}
that the network evolution is rather insensitive to the choice of
$\kappa$, and is not significantly modified even if one sets $\kappa
=0$. In the present simulation we used $\kappa = 0.25$.

\begin{figure}
\begin{center}
\leavevmode\epsfxsize=3.0in\epsfbox{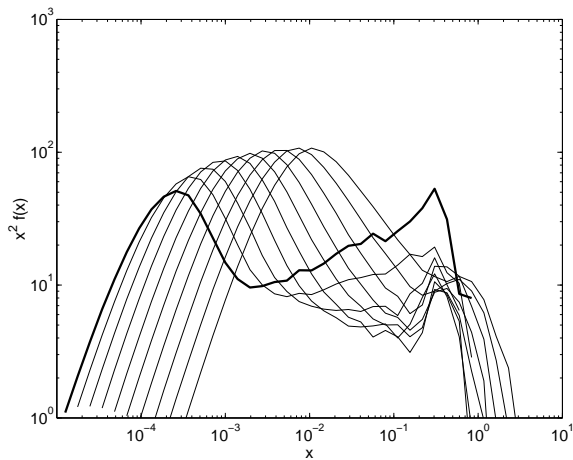}
\end{center}
\caption{The primary loops production function $x^2 f_p(x)$ in equal
logarithmic bins of time centered at times
64, 89, 125, 175, 245, 342, 479, 671, and 939 (bold).}
\label{fig:primary}
\end{figure}

A loop is primary if it is a survivor but none of its ancestors are
survivors.  In Fig.\ \ref{fig:primary}
we plot the primary loop production function $x^2 f_p(x)$, where
$f_p(x) = t^5 n_p(l, t)$ and $n_p$ is the production function of
primary loops with the same conventions as for $n$. The graph thus
shows the fraction of total length produced in primary loops for each
logarithmic interval in $x$.

For a scaling spectrum, we expect $f_p(x)$ to be independent of time.
Instead, we see two peaks closely related to the two peaks in the
power spectrum. The scaling peak (on the right) does not move in $x$
but remains at
\be
	l_p \sim 0.3 t
\ee
and increases in amplitude.
On the other hand, the non-scaling peak (on the left) always remains
at the scale of the initial correlation length, so it moves to smaller
values of $x$, and decreases in amplitude.  At early times the
production of loops is dominated by short lengths, but as long strings
become smoother, the loop production at small scales decreases.  In
contrast, the scaling peak is sub-dominant in the beginning, but as
the time advances it steadily grows.  At even later times, we expect
the peak related to initial conditions to vanish, leaving a scaling
spectrum.

\begin{figure}
\begin{center}
\leavevmode\epsfxsize=3.0in\epsfbox{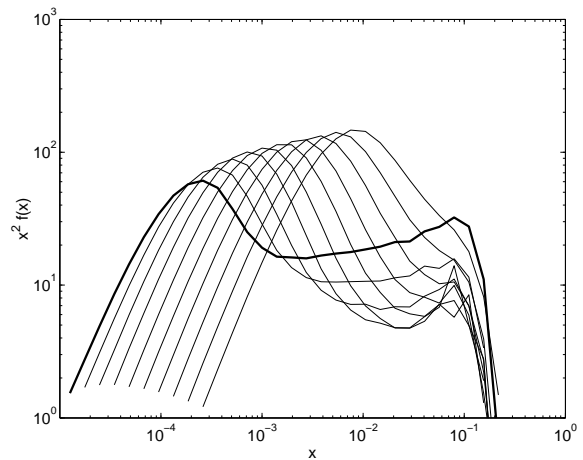}
\end{center}
\caption{The final production function of loops $x^2 f(x)$ for the
same time intervals as Fig.\ \ref{fig:primary}.}
\label{fig:final}
\end{figure}

A primary loop of size $l$ is likely to fragment in time $t \sim l/4$.
The fragmentation process continues until all loops find themselves in
non-self-intersecting trajectories.
In Fig.\ \ref{fig:final}
we plot the final production function of loops $x^2 f(x)$.
As for primary loops, there are two peaks in the final loop production
function: one scaling and one non-scaling. The non-scaling
peak is going down and we expect it eventually to vanish, while the
scaling peak slowly becomes the dominant one and remains at constant
$x$,
\be
	l_f \sim 0.1 t\,.
\ee

\begin{figure}
\begin{center}
\leavevmode\epsfxsize=3.5in\epsfbox{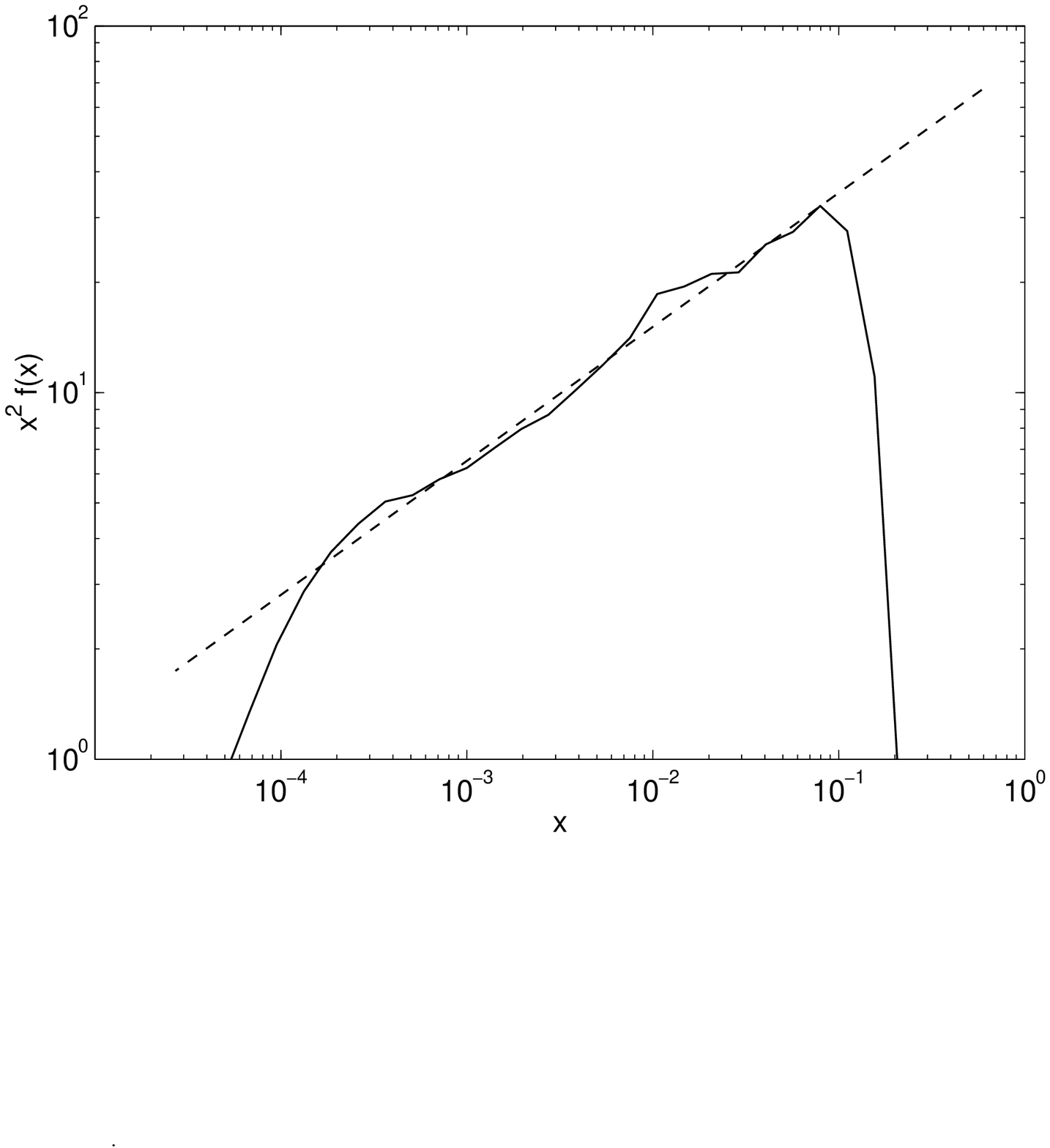}
\end{center}
\caption{The final production function of loops $x^2 f(x)$ from the
last time interval in Fig.\ \ref{fig:final}, considering only primary
loops longer than $8.4$.}
\label{fig:final-big}
\end{figure}

The non-scaling peak in Fig.\ \ref{fig:final} is present only because
of the production of small primary loops shown in Fig.\
\ref{fig:primary}.  If one considers only final loops produced from
primary loops whose length is greater than $8.4$, the small-scale peak is
absent, as shown in Fig.\ \ref{fig:final-big}.  ($l=8.4$ is about
twice the size of the smallest loops in the initial string
network. This particular choice is due to the logarithmic binning we
used for the loop statistics.)

The scaling part of the final loop production function
can be fit by a power law
\be\label{eqn:looppower}
f(x) \approx A x^{-\beta},
\ee
with 
\be\label{beta}
\beta = 1.63\pm 0.03 \,.
\ee
We found $A = 82 \pm 2$ in our simulations, but since a large part of
the loop production is still at small sizes even at the last times
simulated, we expect the final scaling value of $A$ to be somewhat
larger. (The error bars above indicate only statistical variation
between different runs of the simulation.)

The corresponding loop production rate is:
\be\label{eqn:density}
n(l, t) = t^{-5} f(x) \approx A t^{\beta-5} l^{-\beta}\,,
\ee
with a sharp cut-off for large loops at $l \sim 0.1 t$. From
Eq.\ (\ref{beta}), the total number of loops produced is
divergent, while their total length is finite.

Our simulations are done only in flat space, but one can make a
reasonable conjecture about the expanding universe.  The
expansion of the universe stretches the excitations on the strings, so
small-scale structure tends to be reduced relative to the flat-space
case.  Thus we expect the non-scaling peak at the initial condition
size to be eliminated more quickly, and the loop production spectrum
to fall more rapidly toward small sizes in the expanding universe than
in flat space.  If as a result the power index $\beta$ is decreased by
at least 0.63, Eq.~(\ref{eqn:density}) will give a convergent
total number of small loops.

In any case, we expect, on the basis of the simulations described
here, that most of the energy produced in loops from a cosmic string
network appears at scales comparable to the inter-string distance.
This agrees with early expectations \cite{AV81,Kibble85}, but not with
later simulations \cite{BB,AS}.

If there is a scaling process of loop production in an expanding
universe, then loops produced over time at some fixed fraction
$\alpha$ of the horizon size give rise to a spectrum of presently
existing loops \cite{book}
\be\label{eqn:N}
N(l,t) \propto l^{-\beta_c}
\ee
where $\beta_c = 5/2$ for a radiation-dominated universe and $\beta_c
= 2$ for matter.  This spectrum rises more steeply toward
smaller $l$ than the loop production function, Eq.\
(\ref{eqn:looppower}), indicating that late loop production does not
significantly affect the form of the spectrum, Eq.\ (\ref{eqn:N}).
Thus as long as we know that $\beta < \beta_c$ in the expanding
universe, we do not need to know the precise value of $\beta$.  The
form of the loop distribution is given by Eq.\ (\ref{eqn:N}).  This
distribution is cut off at small scales by gravitational back
reaction.

The large value of $\alpha$ suggested by our simulations implies a
higher energy density of the string-generated gravitational wave
background and tighter observational constraints on the string
parameter $G\mu$. The requirement that gravitational waves from
strings do not affect the predictions of big bang nucleosynthesis
can be expressed as \cite{book} (for $\alpha \gtrsim 10^2 G\mu$)
\beq
G\mu\lesssim 10^{-8}\alpha^{-1}.
\label{gw}
\eeq
It follows from the analysis in \cite{Damour} that for $\alpha \gtrsim
10^2 G\mu$ the millisecond pulsar observations yield the bound $G\mu
\lesssim 10^{-7}$, which is similar to (\ref{gw}) for $\alpha\sim 0.1$.
(Both of these bounds assume that the reconnection probability is $p\sim
1$.  We expect $p$ to produce a denser string network
and so make this bound more stringent.  Unfortunately simulation
of small $p$ is difficult because the network evolves much more slowly.)

Recent work \cite{RSB} by Ringeval, Sakellariadou and Bouchet
(RSB) discussed loop production in expanding universe
simulations. They found that the distribution of loops grows steeply
toward small scales, with the power index $\beta\approx 3.0$ in the
radiation era and $\beta\approx 2.5$ in the matter era.  Since these
indices are larger than $\beta_c$, the index $\beta$ in their loop
production function must also have these values, in sharp contrast
with our result, Eq.\ (\ref{beta}). The most likely explanation of
this discrepancy is that the RSB simulation is still in the transient
regime dominated by small loop production. Indeed, the shape of the
loop production function obtained by RSB is similar to that in our
Fig.\ \ref{fig:final} at times $t\lesssim 100$, when the scaling peak
is not yet pronounced.

This work was supported in part by the National Science Foundation
under grants 0353314 and 0457456.

\end{document}